\documentclass[prb,twocolumn]{revtex4-1}  

\usepackage{bbm,graphicx,bm,hyperref}

\def\ii{{\rm i}}

\def\etal#1{#1}

\def\tit#1{}

\def\ket#1{|#1\rangle}

\begin{document}

\title{Magnetization transport in spin ladders and next-nearest-neighbor chains}

\author{Marko \v Znidari\v c}
\affiliation{
Department of Physics, Faculty of Mathematics and Physics, University of Ljubljana, Ljubljana, Slovenia}

\date{\today}

\begin{abstract}
We study magnetization transport at infinite temperature in several spin ladder systems as well as in next-nearest-neighbor coupled spin chains. In the integrable ladder considered we analytically show that the transport is ballistic in sectors with nonzero average magnetization, while numerical simulations of a nonequilibrium stationary setting indicate an anomalous transport in the zero-magnetization sector. For other systems, isotropic Heisenberg ladder and spin chains, showing eigenlevel repulsion typical of quantum chaotic systems, numerical simulations indicate diffusive transport.
\end{abstract}


\maketitle

\section{Introduction}

Understanding transport in quantum and classical systems from first principles has a long history. Perhaps the simplest question one can ask is what is the nature of transport in a given system; is it ballistic, in which case localized disturbances spread with time to a region whose maximal linear size (a diameter) grows linearly with time, or, is it diffusive, in which case the diameter will grow only as a square root of time. In one dimension, being the subject of present work, the situation is clear in systems of non-interacting particles -- in the absence of external scattering effects non-interacting systems are ballistic -- for interacting systems though (also called strongly correlated) such question proves to be very difficult to answer, even in the simplest conceivable models. 

A paradigmatic example of a simple system whose transport properties is difficult to assess is a one-dimensional Heisenberg model~\cite{Heisenberg:28,Dirac:29}. Its anisotropic version (shortly the XXZ chain), with the anisotropy being denoted by $\Delta$, serves as one of the simplest strongly interacting quantum systems. Despite being solvable by the Bethe ansatz~\cite{Bethe:31} its nonequilibrium physics, in particular magnetization transport, is being debated for many years. One can use the so-called Mazur's inequality~\cite{Mazur:69,Zotos:97} to show ballistic transport of energy~\cite{Grabowski:94,Zotos:97} or of magnetization away from the zero-magnetization sector~\cite{Zotos:97} or in the gapless phase $|\Delta| < 1$~\cite{Prosen:11,Ilievski:13}. The main obstacle to a more detailed understanding is the lack of efficient out-of-equilibrium tools, while on the other hand evaluating the linear-response formalism using the Bethe-ansatz solution seems too difficult, except in the simplest case of zero temperature~\cite{Shastry:90}.

One might wonder whether there exists any simple principle that would tell us when to expect diffusion and when not? At first sight an appealing conjecture would be that, due to constants of motion, integrable systems display ballistic transport, while chaotic are diffusive. Unfortunately, there are exceptions to both rules. In the integrable Heisenberg model for $\Delta >1$ and at high temperatures numerics suggests that magnetization transport is diffusive~\cite{Meisner:03,Prelovsek:04,Prosen:09,Steinigeweg:09,Langer:09,Znidaric:11,Jesenko:11,Langer:11,Karrasch:12}, although a more involved picture sometimes emerges~\cite{Steinigeweg:12,Sirker:11}. The same seems to hold also at temperatures below the ground state gap~\cite{Sachdev:97}. The isotropic point $\Delta=1$, being at the transition between the ballistic and the diffusive regime, seems to be even less clear; some numerical investigations suggest anomalous transport~\cite{Znidaric:11,Znidaric:11b,Herbrych:12,Popkov:13}, while others~\cite{Alvarez:02,Fujimoto:03,Meisner:03,Benz:05,Mukerjee:08,Grossjohann:10,Karrasch:12,Karrasch:13} indicate ballistic transport or are inconclusive. In addition, there exists an exactly solvable diffusive (albeit dissipative) 1d model~\cite{Znidaric:10} showing that integrability does not necessarily imply ballistic transport. For chaotic systems things are also not always simple. It has been rigorously shown~\cite{Znidaric:13} that in a special class of XX-type spin ladders (that class, for instance, includes the Hubbard chain) ballistic subspaces exist even-though the model is chaotic. Although probably exceptional, these counterexamples show that the conjecture is not true, at least not in 1d systems. In light of this it is important to gather information on transport in different chaotic and integrable systems. 
 
In the present work we shall study magnetization transport at an infinite temperature and zero average magnetization in a number of spin ladder and next-nearest-neighbor chain systems. Note that next-nearest-neighbor coupled chains can be viewed as ladder systems with a special kind of rung-rung coupling (compare Figs.~\ref{fig:lestve} and~\ref{fig:verige}). Namely, we can in general call a ladder any system that can be viewed  as a nearest-neighbor coupled chain of local $4$-level systems (representing one rung). We should also mention that spinfull 1d chains like, e.g., the 1d Hubbard model~\cite{Shastry:90,finiteL,fujimoto:98,kirchner:99,peres:00,finiteom,Sachdev:97,Prelovsek:04,gu:07,Sirker:11,Prosen:12,Carmelo:12}, can be, via Jordan-Wigner transformation, rewritten as spin ladder models. Spin ladder systems are not just of theoretical interest but are realized in a number of compounds, for a review see Refs.~\onlinecite{review,Dagotto:99}.

Apart from one integrable ladder, we shall exclusively focus on systems with strong chaos. In the integrable SU(4) ladder we numerically find anomalous transport in a subsector with zero-magnetization and analytically prove ballistic transport in sectors with non-zero magnetization. For all other models studied (isotropic Heisenberg ladder and XX chain as well as Heisenberg chain with next-nearest-neighbor coupling), all being quantum chaotic, we numerically find diffusive or very-close-to diffusive transport.

\section{Methods}
There are different ways to numerically assess quantum transport. One is via linear response theory by evaluating the equilibrium time-dependent current autocorrelation function. In numerical calculations one is always limited to finite-size systems causing two effects: for finite $L$ and time $t$ the correlation function $C(t)$ might not yet converge to its thermodynamic limit value and, going with $t\to \infty$ the correlation function will not decay to zero, even in a diffusive system, but will rather have finite-size fluctuations. Therefore, due to these finite-$L$ and finite-$t$ effects great care is needed to correctly evaluate the limits $\lim_{t\to \infty}\lim_{L \to \infty}$ in the correct order. Another way of studying transport is to directly simulate nonequilibrium states. There are two possibilities, one can study the transient dynamics of initial nonequilibrium states like, e.g., spreading of localized packets and calculating how fast their width increases with time, or, one can go to a stationary setting in which constant driving is applied to a system. The latter approach has the advantage that there are no finite-time effects, only finite-size, as one, by definition, studies a nonequilibrium stationary state reached after an infinite time. In the present study we shall use a nonequilibrium stationary setting.

The following subsections describe the methods used and do not present any new material. Sec.~\ref{sec:Lin} describes the Lindblad formalism and presents the Lindblad operators used for simulations; in Sec.~\ref{sec:num} some details are given about numerical simulations, in Sec.~\ref{sec:anomal} we repeat basic notions about normal and anomalous transport, while in Sec.~\ref{sec:chaos} we present the level spacing criterion of quantum chaos.

\subsection{Lindblad master equation}
\label{sec:Lin}
A nonequilibrium situation will be induced by a boundary coupling to magnetization reservoirs. These can, with certain probabilities, flip the boundary spin either up or down. If these probabilities are different at two ladder ends the driving will cause a nonequilibrium situation. Spin flips at the boundary are described in an effective way with the so-called Lindblad operators $L_k$, while the density matrix describing the ladder evolves according to the Lindblad master equation,
\begin{eqnarray}
\label{eq:Lin}
{{\rm d}}\rho(t)/{{\rm d}t}=\ii [ \rho(t),H ]+ {\cal L}^{\rm dis}(\rho(t))={\cal L}(\rho(t)),\\
 {\cal L}^{\rm dis}(\rho(t))=\sum_k [ L_k \rho(t),L_k^\dagger ]+[ L_k,\rho(t) L_k^{\dagger} ]. \nonumber
\end{eqnarray}
Provided the dissipative part ${\cal L}^{\rm dis}$ is nonzero one will typically have a single stationary state $\rho_\infty$, being the solution of ${\cal L}(\rho_\infty)=0$, to which an arbitrary initial state $\rho(0)$ converges after a long time, $\rho_\infty = \lim_{t\to\infty}\rho(t)$. In a nonequilibrium setting such a state $\rho_\infty$ is called the nonequilibrium steady state (NESS). The summation over $k$ in (\ref{eq:Lin}) goes over all Lindblad operators. What kind of Lindblad operators are used depends on each specific situation.

Before specifying in detail the Lindblad operators used, let us comment on the applicability of the Lindblad equation within the context of quantum transport. The Lindblad equation can be derived from microscopic equations of motion of the system plus reservoirs under certain, from the condensed-matter perspective, rather restrictive conditions of a weak coupling and a fast decaying environmental correlations~\cite{Breuer:02}. While these conditions are sometimes well satisfied, e.g., in quantum optical systems where the environment is fast, this is not so in condensed matter. Environmental degrees there (electrons in the leads, phonons,...) are not necessarily fast compared to the timescale of the system of interest. As a consequence, the evolution equation for the system will not be local in time, like the Lindblad equation (\ref{eq:Lin}), but will in general be non-local with a nontrivial integral kernel accounting for memory effects. While memory-effects can play a role in a transient finite-time behavior they are not expected to be important in the long time limit of nonequilibrium stationary states considered here. In certain situations one can even show exactly that the memory effects (i.e., non-Markovian effects) play no role for the NESS~\cite{Jesenko:13}.

The Lindblad operators modeling the reservoirs will differ depending on whether we want to study ladders or next-nearest-neighbor chains (which can be viewed as ladders with a diagonal inter-rung coupling, see Fig.~\ref{fig:verige}). For ladders both spins in the first and in the last rung are coupled to the reservoir. The eight Lindblad operators that we use are
\begin{eqnarray}
L_{1}&=&\sqrt{\Gamma(1- \mu)}\,\sigma^{+}_1, \quad L_{2}=\sqrt{\Gamma(1+ \mu)}\,\sigma^{-}_1, \nonumber \\
L_{3}&=&\sqrt{\Gamma(1+ \mu)}\,\sigma^{+}_L, \quad L_{4}=\sqrt{\Gamma(1- \mu)}\,\sigma^{-}_L, \nonumber \\
L_{5}&=&\sqrt{\Gamma(1- \mu)}\,\tau^{+}_1, \quad  L_{6}=\sqrt{\Gamma(1+ \mu)}\,\tau^{-}_1, \nonumber \\
L_{7}&=&\sqrt{\Gamma(1+ \mu)}\,\tau^{+}_L, \quad  L_{8}=\sqrt{\Gamma(1- \mu)}\,\tau^{-}_L,
\label{eq:ladderL}
\end{eqnarray}
where $\sigma^\alpha_k$ and $\tau^\alpha_k$ are Pauli matrices on the 1st and the 2nd ladder leg, respectively, and $\sigma^\pm=(\sigma^{\rm x} \pm {\rm i}\, \sigma^{\rm y})/2$, $\tau^\pm=(\tau^{\rm x} \pm {\rm i}\, \tau^{\rm y})/2$. $L$ is the number of rungs. For n.n.n coupled chains only the left-most and the right-most spins are coupled to reservoirs. The four Lindblad operators are in this case
\begin{eqnarray}
L_{1}&=&\sqrt{\Gamma} \sqrt{1-\mu}\sigma^+_1,\quad L_{2}=\sqrt{\Gamma} {\sqrt{1+\mu}} \sigma^-_1, \nonumber \\
L_{3}&=&\sqrt{\Gamma} \sqrt{1+\mu}\sigma^+_L,\quad L_{4}=\sqrt{\Gamma} {\sqrt{1-\mu}} \sigma^-_L.
\label{eq:chainL}
\end{eqnarray}
For chains $L$ is the chain length. The coupling $\Gamma$ in both cases plays no essential role and we fix it to $\Gamma=1$. All NESS states obtained with such Lindblad operators studied here are unique.

Note that the precise form of Lindblad operators, and their number, is not expected to play any role on the results presented, as long as they induce a NESS at an infinite temperature. In quantum chaotic systems the value of the diffusion constant is also not influenced by the choice of Lindblad operators. Provided the boundary effects are small, which is the case at high temperature~\cite{Znidaric:10b}, and for non-ballistic systems, the bulk properties should be independent of the details of driving. For ladders, being symmetric with respect to the exchange of two legs (see Fig.~\ref{fig:lestve}), the natural choice is 8 Lindblad operators (\ref{eq:ladderL}), while for n.n.n. coupled chains, without that symmetry (see Fig.~\ref{fig:verige}), the natural choice is 4 Lindblad operators (\ref{eq:chainL}). The choice used in the present work is perhaps the simplest because it induces states at an infinite temperature and has been used in a number of our previous studies, see also, e.g. Ref.~\onlinecite{Michel:08}. 

The most important parameter in Lindblad operators is the driving strength $\mu$. For zero driving $\mu=0$ and all systems studied one can easily show (see, e.g., Ref.~\onlinecite{Znidaric:13}) that the NESS is a trivial $\rho_\infty \sim \mathbbm{1}$, that is, it is an equilibrium state at an infinite temperature. For nonzero $\mu$ the stationary state is a true nonequilibrium state with a nonzero current flowing through the system. For sufficiently small $\mu$ the NESS is still close to an identity density matrix and one can expand it in a series over $\mu$, $\rho_\infty \propto \mathbbm{1}+\mu\, A + {\cal O}(\mu^2)$. Although for non-solvable systems the precise form of $A$ can not be explicitly calculated in the thermodynamic limit (see though Ref.~\onlinecite{Znidaric:10} for a solvable case where it can), one can nevertheless make some useful general statements. For the driving used, Eqs.~\ref{eq:ladderL} or~\ref{eq:chainL}, the operator $A$ contains, among other, also local current and magnetization operators. This means that for small $\mu$ the expectation values of magnetization density and current are trivially proportional to $\mu$. The fact that for small $\mu$ the NESS is close to an identity also has consequences for the temperature of these NESSs. In general, provided that the nonequilibrium is locally sufficiently weak (e.g., taking $L\to \infty$ at fixed driving strength) one can determine the local temperature and chemical potential by comparing the expectation values of local operators in the NESS with the expectation values in an equilibrium grand-canonical state at a given temperature and chemical potential, see Ref.~\onlinecite{Znidaric:10b}. However, for the driving used here (Eq.~\ref{eq:ladderL} or~\ref{eq:chainL}) and small $\mu$ the situation is much simpler. Namely, because the NESS is close to $\mathbbm{1}$ we immediately know that the expectation value of the energy density will also be close to zero and that such states are close to an infinite temperature. Therefore, we are studying nonequilibrium systems at an infinite temperature~\cite{footE}. The driving used is also symmetric with respect to the left/right end and the NESS obtained has always zero expectation value of the total magnetization (i.e., in the fermionic language this would be called a half-filling). 

The ladder and chain systems that shall be considered (without driving) all conserve the total magnetization in the $z$ direction. The corresponding unitary symmetry is $U=\exp{(-\ii\alpha \sum_j \sigma_j^{\rm z})}$, with $UHU^\dagger=H$. Because the dissipative Lindblad term ${\cal L}^{\rm dis}$ (\ref{eq:ladderL},\ref{eq:chainL}) is also invariant under such $U$ (this is a consequence of $U\sigma^+U^\dagger={\rm e}^{-\ii 2\alpha }\sigma^+$ and the fact that ${\cal L}^{\rm dis}$ does not depend in the phase of the Lindblad operators), where the invariance for ${\cal L}^{\rm dis}$ means that $U {\cal L}^{\rm dis}(\rho)U^\dagger={\cal L}^{\rm dis}(U\rho U^\dagger)$, nonequilibrium steady states considered in the present work are all independent of the optional homogeneous magnetic field in the $z$ direction added to $H$. That is, if $\rho_\infty$ is the NESS for ${\cal L}$ with $H$, then the same $\rho_\infty$ is the NESS also for ${\cal L'}$ with $H'=H+B \sum_j \sigma_j^{\rm z}$. This is a general consequence of the symmetry of the master equation. The proof is very simple. Let us denote by $V$ a general unitary symmetry, and by $C$ the corresponding conserved quantity. Let $V$ be an exact symmetry of the Liouvillian (\ref{eq:Lin}), that is $V {\cal L}(\rho)V^\dagger={\cal L}(V\rho V^\dagger)$. Provided the NESS is unique (with our driving this is always the case) it must be invariant under $V$, meaning that $V\rho_\infty V^\dagger=\rho_\infty$. This can be seen by noting that $V {\cal L}(\rho_\infty)V^\dagger=0={\cal L}(V\rho_\infty V^\dagger)$, see also e.g., Ref.~\onlinecite{Popkov:13b}. This means that in the eigenbasis of a corresponding conserved quantity $C$ matrix $\rho_\infty$ is block-diagonal, while matrix $C$ is diagonal with identical elements within each diagonal block. $C$ and $\rho_\infty$ therefore commute and, if $\rho_\infty$ satisfies $\ii [\rho_\infty,H]+{\cal L}^{\rm dis}(\rho_\infty)=0$, it also satisfies $\ii [\rho_\infty,H+C]+{\cal L}^{\rm dis}(\rho_\infty)=0$, i.e., $\rho_\infty$ is also the NESS state for $H'=H+C$.

\subsection{Numerical method}
\label{sec:num}
Because we want to study the system's properties in a stationary state we have to obtain $\rho_\infty$. There are two possibilities: one can either solve the stationary equation ${\cal L}(\rho_\infty)=0$, or, one can integrate the Lindblad equation (\ref{eq:Lin}), obtaining $\rho(t)$ and from it the NESS in the limit $t \to \infty$. We use the latter method by first writing $\rho(t)$ in a matrix product form with matrices $A_k^s$ of fixed dimension $M$, describing a site $k$ and an element $s$ of a local operator basis. Ladders as well as n.n.n. chains are treated as a ladder system with an arbitrary coupling between two nearest-neighbor rungs. One rung is considered as a single site $k$, so that the dimension of the operator basis at one site is $4^2$ (i.e., the number of different values of the index $s$ in matrices $A_k^s$). The total number of complex parameters describing a state $\rho(t)$ of a ladder with $L$ rungs is therefore $16LM^2$. Choosing a large enough $M$ any state $\rho(t)$ can be written in such a matrix product operator form. Time evolution is then evaluated using the time-dependent density renormalization group method~\cite{Vidal:03} (time-evolved block decimation, TEBD) by writing a short-time propagator $e^{{\cal L}\, \Delta t}$ as a series of single and two-site transformations. The method we use is an adaptation~\cite{Prosen:09} for dissipative systems in which the optimality of a matrix product decomposition is preserved by reorthogonalizations, for details see Ref.~\onlinecite{Znidaric:10c}. Evaluating two-site transformations exactly would lead to an exponentially increasing (in time) matrix dimension $M$. Numerically this can not be handled and one truncates dimension after each transformation to a fixed size $M$. This truncation is the main source of errors in the numerical method. How large should $M$ be depends on the amount of entanglement that a state $\rho(t)$ has in the operator space. For instance, the equilibrium state at an infinite temperature is a product state (a product of identities at each site) and can be represented by matrices of size $M=1$. For small $\mu$, where $\rho(t)$ is still close to $\mathbbm{1}$, we therefore expect that one can do with a reasonably small $M$. This is the reason why simulations at an infinite temperature require the smallest $M$ and are therefore the easiest~\cite{Znidaric:08}. In our simulations we used matrix sizes of up-to $M=150$ and ladder lengths of $L\le 100$. Because the costliest operation in the algorithm is a singular value decomposition of a matrix of size $d\,M$, if $d$ is the dimension of local operator space, the time needed for one time-step scales as $\sim M^3$ and quickly becomes unmanageable for larger $M$. We typically performed simulations at an increasing values of $M$ and observed the convergence of e.g. the current. We deemed results as having converged if the difference in currents between the two largest $M$ was less than $\sim 1-2\%$ (for the hardest integrable ladder see also~\cite{foot1}). Note also that, because $d=4^k$ for a ``ladder'' with $k$ legs, adding one leg increases the computational time by a factor of $64$ (keeping $M$ and $L$ the same). Simulating ladders with two legs is therefore about $64$ times more time consuming than simulating chains. Therefore, beyond 2-leg ladders, simulations soon become too time-consuming. However, one can expect on general grounds that the transport will be typically diffusive in systems with more legs as integrable cases are rarer in higher dimensional systems. 

\subsection{Assessing transport}
\label{sec:anomal}
Once the NESS is obtained -- after time $t$ that is given by the inverse of the Liouvillean gap -- the expectation values of any local operator can be evaluated. Practically, the simulation is run until the current converged to a time-independent value, which typically happened after a time that was some multiple of $L$. Our main focus is on the magnetization current and on the magnetization profile along the ladder/chain. Fixing the driving $\mu$, typically at~\cite{foot0} $\mu=0.2$, we study how the magnetization current $j$ scales with the system length $L$. If the scaling is $j \sim 1/L$, the system is diffusive and obeys a phenomenological transport law
\begin{equation}
j = -D \nabla z,
\label{eq:Four}
\end{equation}
where $\nabla z$ is the magnetization gradient and $D$ is the size-independent transport coefficient (diffusion constant). Other extreme situation would be when $j$ is independent of $L$, signaling ballistic transport. Transport that is intermediate between ballistic and diffusive is called anomalous~\cite{Dhar} with a scaling $j \sim 1/L^\alpha$ with $0<\alpha<1$. The nature of transport can be also inferred from the magnetization profile. For small magnetization the profile is linear for diffusive systems while it is non-linear in the case of an anomalous transport where $D$ can be considered to be length-dependent. 

All the above statements about the scaling should be investigated in the thermodynamic limit $L \to \infty$. Even though we limit ourselves, besides one integrable case, to fully chaotic systems, where the convergence with $L$ is expected to be the fastest, it turns out that in some cases sizes $L \sim 100$, though rather large for a quasi-exact simulation of a strongly interacting quantum system, still might not be large enough to reach the thermodynamic limit. Going to significantly larger sizes is at present not possible due to the rapidly increasing simulation times. The simulation time increases with $L$ because the convergence time to $\rho_\infty$ increases, but even more significantly, the required matrix size $M$ also increases because the observables, like the current, decrease with $L$ and so a larger $M$ is required to obtain the same relative accuracy in $j$.

\subsection{Checking for quantum chaos}
\label{sec:chaos}
Despite the exceptions~\cite{Znidaric:13}, one in general expects that for non-integrable strongly interacting quantum system, in other words for systems displaying characteristic features of quantum chaos, transport is diffusive. In order to convince ourselves that the systems we study are indeed not being close to integrability, we have checked their chaoticity by calculating the spacing distribution of nearest energy levels. The so-called level spacing distribution $p(s)$ is a standard criterion of quantum chaos in Hamiltonian systems~\cite{Haake:10}. In chaotic systems there are no selection rules and the eigenenergies will ``repel'' each other, leading to a deficit of small spacings between two consecutive eigenenergies. In quantum chaotic systems with time-reversal symmetry the expected theoretical level spacing distribution is well described by the so-called Wigner's surmise for the orthogonal ensemble,
\begin{equation}
p(s)=s\frac{\pi}{2} \exp{(-s^2\pi/4)}.
\label{eq:Wigner}
\end{equation}
In integrable system there are selection rules due to constants of motion, resulting in an exponential form of $p(s)=\exp{(-s)}$. One should bear in mind that in order to see chaotic level statistics (\ref{eq:Wigner}) explicit symmetries of a system have to be taken into account. Spacing has to be calculated within a single symmetry class. The symmetries of the systems studied are described in the Appendix~\ref{app:symm}. We always use open boundary conditions, which though has no effect on the level spacing distribution in chaotic systems.

\section{Results for ladders}

There have been a number of works studying magnetization transport~\cite{Sachdev:97,Meisner:03,Langer:09,Znidaric:13} as well as heat transport~\cite{Alvarez:02b,Orignac:03,Zotos:04,Boulat:07} in spin ladders. The prevailing conclusion is that in non-integrable ladders at high temperatures transport is diffusive, being in-line with the general rule of expecting diffusion in non-integrable systems. Numerical studies have been mostly limited to systems with less than $L=20$ rungs; in the present work we shall study significantly larger systems. For a review on ladder systems, including references to an extensive experimental work, see Refs.~\onlinecite{review,Dagotto:99}. For studies of transport in the 1d Hubbard model, that can be equivalently rewritten as a spin ladder, see Refs.~\onlinecite{Shastry:90,finiteL,fujimoto:98,kirchner:99,peres:00,finiteom,Sachdev:97,Prelovsek:04,gu:07,Sirker:11,Prosen:12,Carmelo:12}.

We shall name different ladder systems according to the type of the coupling between nearest-neighbor sites. The XX-type is a coupling of the form $\sigma_i^{\rm x} \sigma_{i+1}^{\rm x}+\sigma_i^{\rm y} \sigma_{i+1}^{\rm y}$, the XXZ-type is a coupling of the form $\sigma_i^{\rm x} \sigma_{i+1}^{\rm x}+\sigma_i^{\rm y} \sigma_{i+1}^{\rm y}+\Delta \sigma_i^{\rm z} \sigma_{i+1}^{\rm z}$, while an isotropic Heisenberg coupling is equal to the XXZ coupling with the anisotropy $\Delta=1$, i.e., an XXX coupling.

\begin{figure}[ht!]
\centerline{\includegraphics[width=0.25\textwidth]{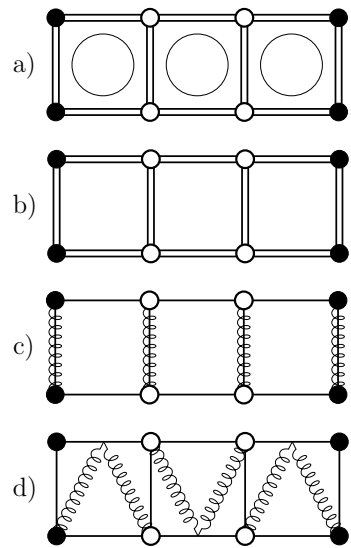}}
\caption{Schematic representation of different spin ladders: a) integrable ladder, Eq.~(\ref{eq:Betheladder}), b) isotropic Heisenberg ladder, Eq.~(\ref{eq:Heisladder}), c) XX-ladder, d) free fermions on a ladder. A straight line denotes an XX-type coupling, a spring a ZZ-type coupling, a double line is an isotropic Heisenberg coupling while a straight line with two springs in d) is a coupling involving 4 sites (see text). Full points mark sites that are coupled to a reservoir described by Eq.~(\ref{eq:ladderL}).} 
\label{fig:lestve}
\end{figure}

Ladder systems that shall be studied are depicted in Fig.~\ref{fig:lestve}a and Fig.~\ref{fig:lestve}b. The same methodology that we use has been used before to study the so-called XX ladder, Fig.~\ref{fig:lestve}c, with an XX-type coupling along the legs and an XXZ-type in the rungs. A special case of such an XX ladder is the 1d Hubbard model obtained if the coupling in rungs is $\sigma_i^{\rm z} \tau_{i}^{\rm z}$. As shown in Ref.~\onlinecite{Prosen:12} the 1d Hubbard model is diffusive under symmetric driving at infinite temperature. This diffusive transport is not changed in the presence of an additional XX-type coupling in rungs~\cite{Znidaric:13}. It is instructive to rewrite a tight-binding system of free fermions on a ladder in spin language. Namely, for a tight-binding model we know that is is ballistic because it is equivalent to a system of free fermions. Using the Jordan-Wigner transformation it can be written as the ladder shown in Fig.~\ref{fig:lestve}d with the 4-site coupling interchangeably connecting neighbors in the upper/lower leg being of the form $(\sigma_k^{\rm x}\sigma_{k+1}^{\rm x}+\sigma_k^{\rm y}\sigma_{k+1}^{\rm y})\tau_k^{\rm z}\tau_{k+1}^{\rm z}$ (written here for the upper leg). We have numerically checked (data not shown) that such a ladder coupling indeed results in a ballistic magnetization transport. Note that the $\tau_k^{\rm z}\tau_{k+1}^{\rm z}$ term in the above coupling is absolutely crucial for the ballistic transport to appear; without it one would have an ordinary XX-type ladder displaying diffusive transport~\cite{Znidaric:13}.

\subsection{Isotropic Heisenberg ladder}
\begin{figure}[ht!]
\centerline{\includegraphics[angle=0,width=0.35\textwidth]{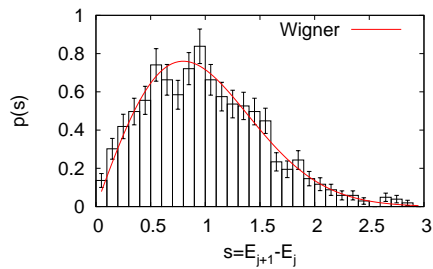}}
\caption{Level spacing distribution for the isotropic Heisenberg ladder (\ref{eq:Heisladder}). Parameters are $U=1$, $L=8$ and four symmetry sectors with $Z=0$ from the subspace with zero total spin are used (1026 spacings in total), see the Appendix~\ref{app:symm} for details about symmetries. Error bars denote one standard deviation obtained from the square-root of the number of spacings in a given bin.
}
\label{fig:lsdIsoHeis}
\end{figure}
The isotropic Heisenberg ladder is described by
\begin{equation}
H=\sum_{i=1}^{L-1} \bm{\sigma}_i\cdot \bm{\sigma}_{i+1}+\bm{\tau}_i\cdot \bm{\tau}_{i+1}+U\,\sum_{i=1}^{L} \bm{\sigma}_i\cdot \bm{\tau}_i.
\label{eq:Heisladder}
\end{equation}
\begin{figure}[ht!]
\centerline{\includegraphics[angle=-90,width=0.45\textwidth]{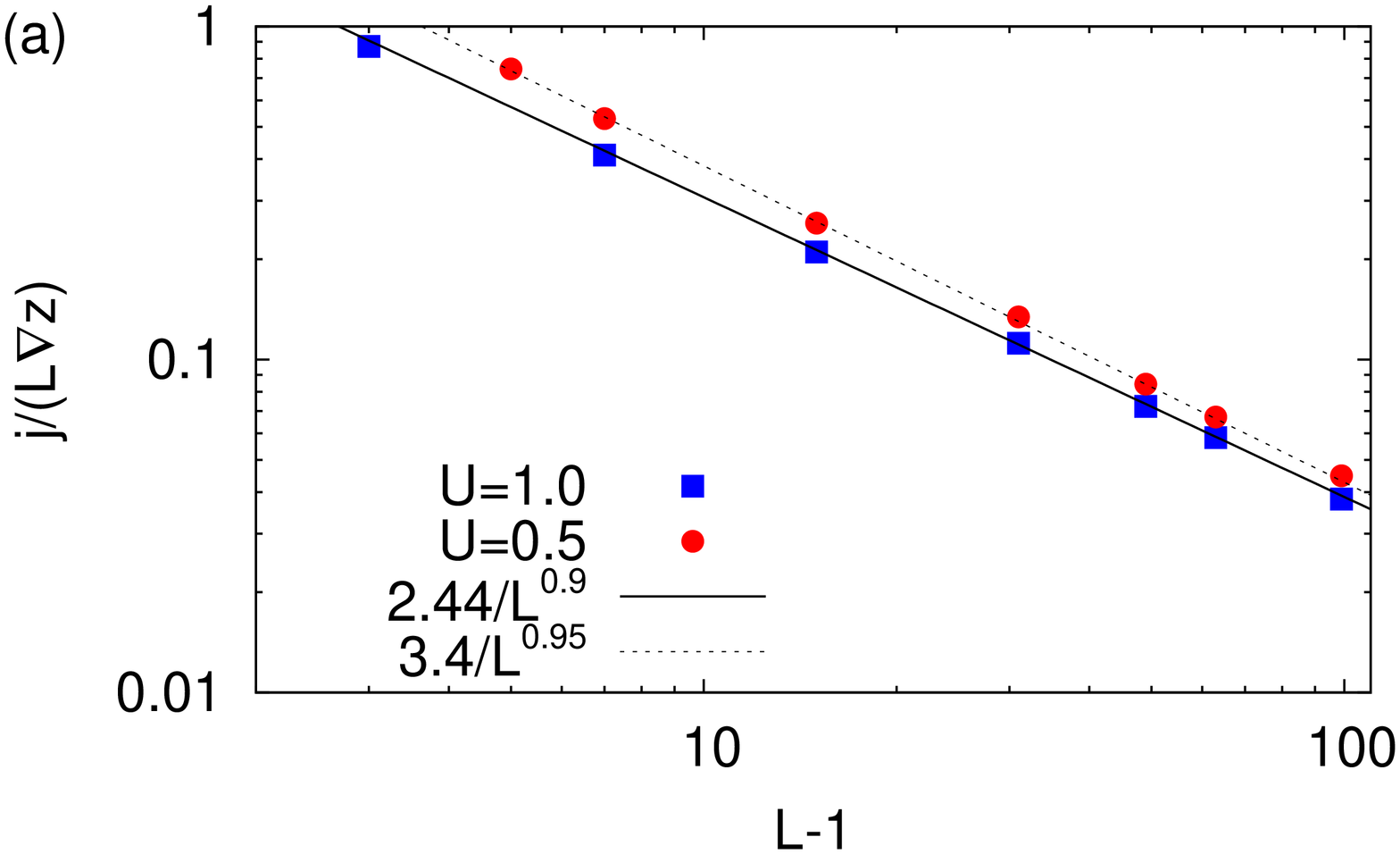}}
\centerline{\includegraphics[angle=-90,width=0.45\textwidth]{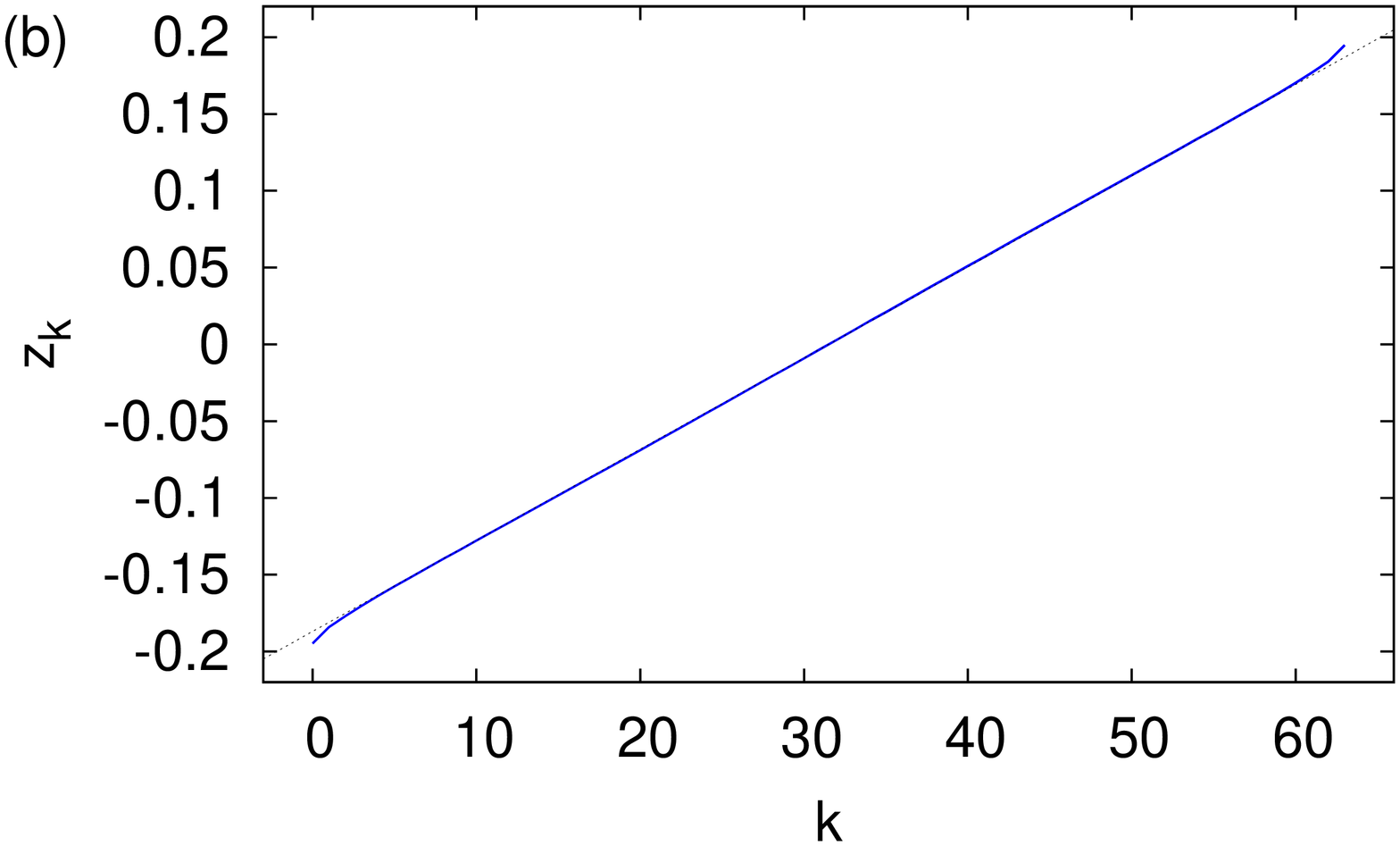}}
\caption{a) Dependence of the scaled magnetization current on $L$, b) magnetization profile in one of the legs ($U=1, L=64$). All is for the isotropic Heisenberg ladder (\ref{eq:Heisladder}) with driving $\mu=0.2$; in a) data is shown for $U=1.0$ and $U=0.5$.}
\label{fig:jh}
\end{figure}
The isotropic Heisenberg ladder, and in particular its version with different coupling strengths along rungs and legs ($U \neq 1$), is realized in some materials and is therefore also experimentally relevant model~\cite{review,Dagotto:99}. System (\ref{eq:Heisladder}) has a nonzero spin gap~\cite{Barnes:93}. It is quantum chaotic as is indicated by the good agreement of the level spacing distribution with the Wigner's surmise (\ref{eq:Wigner}) demonstrated in Fig.~\ref{fig:lsdIsoHeis}. Regarding magnetization transport, in Ref.~\onlinecite{Langer:09} it has been found that initial localized packets spread out diffusively at zero temperature. 

In our stationary nonequilibrium setting we use a symmetric driving of Eq.~(\ref{eq:ladderL}) so that in the NESS $\rho_\infty$ magnetization flows only along both legs while there is no current in the rungs. The driving is chosen to be $\mu=0.2$ for which we are still in the linear response regime. It has been explicitly checked that for $\mu=0.1$ the results in Fig.~\ref{fig:jh}a would be almost indistinguishable from the presented ones for $\mu=0.2$, thereby confirming the validity of the linear response. Note that for very small driving $\mu$ the expectation values of current and magnetization are trivially proportional to $\mu$. The current operator is defined via a continuity equation for local magnetization $\sigma_k^{\rm z}+\tau_k^{\rm z}$, resulting in $j^{\rm tot}_k={\rm i}[\sigma_k^{\rm z}+\tau_k^{\rm z},h_{k,k+1}]$, where $h_{k,k+1}$ is the local hamiltonian density. For the model in Eq.(\ref{eq:Heisladder}) we obtain $j_k^{\rm tot}=j_k^{\sigma}+j_k^\tau$, with the current operator in the upper leg $j^\sigma_k=2(\sigma_k^{\rm x} \sigma_{k+1}^{\rm y}-\sigma_k^{\rm y} \sigma_{k+1}^{\rm x})$, and a similar expression for $j_k^\tau$ in the lower leg. Due to the symmetric driving of both legs (\ref{eq:ladderL}) in the NESS both currents are the same and, due to continuity, independent of the site $k$. We shall therefore simply study the current in one of the legs and denote $j=\langle j_k^\sigma \rangle=\langle j_k^\tau \rangle$, with the averages being expectation values in the NESS $\rho_\infty$. In Fig.~\ref{fig:jh} we show the current and magnetization profile $z_k={\rm tr}\,(\sigma_k^{\rm z} \rho_\infty)$ in the NESS. While the magnetization profile is linear (with only very small deviations at few edge sites), suggesting diffusion, the scaling of the current shows small deviation from a diffusive $\sim 1/L$. Observe that if one plots the scaled current $j/(L\nabla z)$ vs. the system size $L$, as is the case in all our plots (e.g., Fig.~\ref{fig:jh}(a)), then the prefactor in front of the $1/L$ scaling (i.e., the slope in a log-log plot) is equal to the diffusion constant. At $U=1.0$ the scaling is $j \sim 1/L^{0.9}$, while at $U=0.5$ it is $j \sim 1/L^{0.95}$. Note that at $U=0$ one would have two uncoupled isotropic Heisenberg chains for which an anomalous $j \sim 1/L^{0.5}$ scaling has been observed~\cite{Znidaric:11,Znidaric:11b}. From the finite-size data presented it is difficult to make a definite conclusion whether magnetization transport in the Heisenberg ladder is diffusive or not in the thermodynamic limit. Considering the rather linear magnetization profiles we deem it plausible that the small deviations observed are due to finite-size effect and the transport would become diffusive in the thermodynamic limit.

\subsection{Integrable ladder}
The Hamiltonian is
\begin{equation}
H=\sum_{j=1}^{L-1} (1+\bm{\sigma}_j\cdot \bm{\sigma}_{j+1})(1+\bm{\tau}_j\cdot \bm{\tau}_{j+1})+4U\sum_{j=1}^{L} \bm{\sigma}_j\cdot \bm{\tau}_j.
\label{eq:Betheladder}
\end{equation}
It is a Heisenberg ladder with an additional four-spin interaction~\cite{Nersesyan:97}. At $U=0$ the model is called the spin-orbital model~\cite{Li:98} and can be obtained as the large-$U$ limit of the two-orbital Hubbard model at quarter filling~\cite{Kugel:73}. The spin-orbital model can be, up-to an irrelevant constant, written as $H_{\rm SU(4)}=H(U=0)=\sum_j P_{j,j+1}$, where $P_{j,j+1}$ is the permutation operator on two rungs, $P_{j,j+1} \ket{\alpha,\beta}=\ket{\beta,\alpha}$, and $\ket{\alpha},\ket{\beta}$ are two arbitrary rung states. Alternatively, it can be expressed in terms of generators $G^{\nu,\lambda}=\ket{\nu}\langle \lambda |$ of the SU(4) group, $H_{\rm SU(4)} \sim \sum_k \sum_{\nu,\lambda} G_k^{\nu,\lambda} G_{k+1}^{\lambda,\nu}$. The interaction is therefore SU(4) invariant and the spin-orbital model can be considered to be a generalization of the isotropic Heisenberg chain (that has an SU(2) symmetry) and is sometimes called the SU(4) Heisenberg model. It is Bethe ansatz solvable in one dimension by the general method~\cite{Sutherland:75} for systems with permutation interaction. The one-rung operators $C_1=\sum_j \sigma_j^{\rm z}+\tau_j^{\rm z}$, $C_2=\sum_j \sigma_j^{\rm z}\tau_j^{\rm z}$ and $C_3=\sum_j \sigma_j^{\rm x}\tau_j^{\rm x}+\sigma_j^{\rm y}\tau_j^{\rm y}+\sigma_j^{\rm z}\tau_j^{\rm z}$ are conserved quantities for any $U$, while at $U=0$ also all three components of $\sum_j \bm{\sigma}_j$ and $\sum_j \bm{\tau}_j$ are conserved. The nonzero rung interaction $U$ (\ref{eq:Betheladder}), being equal to $C_3$, plays the role of a chemical potential, preserving integrability of the system~\cite{Wang:99} for any $U$. The model is gapless~\cite{Wang:99} for $U<1$ and gapped for $U>1$. 

\subsubsection{Ballistic transport in nonzero-magnetization sectors}
Let us for a moment consider the $H_{\rm SU(4)}$ obtained for $U=0$. Because the Hamiltonian is the sum of nearest-neighbor transpositions one can easily construct invariant subspaces that will display ballistic transport. Taking the singlet $\ket{S}=(\ket{01}-\ket{10})/\sqrt{2}$ and triplet states $\ket{T}=(\ket{01}+\ket{10})/\sqrt{2}, \ket{O}=\ket{00}, \ket{I}=\ket{11}$ for the rung basis, and for instance the initial state of the ladder $\ket{S\ldots SIIIS\ldots S}$, we can see that $H_{\rm SU(4)}$ acting on such a state will cause the left-most and the right-most $I$s to spread ballistically to the left and right, respectively, causing two ballistic fronts. Although such construction is similar to the one in Ref.~\onlinecite{Znidaric:13}, the two situations are fundamentally different. $H_{\rm SU(4)}$ is integrable, and, as we shall show, the energy current is a constant of motion causing ballistic transport in sectors with nonzero magnetization, whereas the XX ladder discussed in Ref.~\onlinecite{Znidaric:13} is quantum chaotic with a more complicated dynamics than just transpositions (for instance, there the energy current is not a constant of motion). To see why model (\ref{eq:Betheladder}) is ballistic away from a zero-magnetization sector let us first define currents. The magnetization current operator is independent of $U$ and is $j_k^{\rm tot}=j_k^\sigma(1+\bm{\tau}_k\cdot \bm{\tau}_{k+1})+j_k^\tau(1+\bm{\sigma}_k \cdot \bm{\sigma}_{k+1})$, where $j_k^{\sigma,\tau}$ are the same chain currents as for the isotropic Heisenberg ladder. The local energy current, defined by $j^{\rm E}_k=\ii [h_{k-1,k},h_{k,k+1}]$, is the sum of an $U$-independent term and a term proportional to $U$, $j^{\rm E}_k=j^{\rm E}_k(U=0)+U\cdot j_k^{\rm E}(U\neq0)$. $j_k^{\rm E}(U=0)$ is simply the energy current of the $H_{\rm SU(4)}$ model and is (written for $k=2$)
\begin{eqnarray}
j_2^{\rm E}(U=0)=[ &&(\sigma_1^{\rm z} j^\sigma_{23}+\sigma^{\rm z}_2 j^\sigma_{31}+\sigma^{\rm z}_3 j^\sigma_{12})\\
&&(\mathbbm{1}+h^\tau_{12}+h^{\tau}_{13}+h^\tau_{23}) + (\sigma \leftrightarrow \tau )], \nonumber
\end{eqnarray}
with $j^\sigma_{jk}=2(\sigma_j^{\rm x}\sigma_k^{\rm y}-\sigma_j^{\rm y}\sigma_k^{\rm x})$ and $h^\tau_{jk}=\bm{\tau}_j \cdot \bm{\tau}_k$, and $(\sigma \leftrightarrow \tau)$ meaning all preceding terms with $\sigma$ and $\tau$ matrices being interchanged. One can show that, taking periodic boundary conditions, the total energy current of the $H_{\rm SU(4)}$ model, $J^{\rm E}_0=\sum_k j_k^{\rm E}(U=0)$, is an exact constant of motion, $[J^{\rm E}_0,H]=0$, regardless of $U$ (note that $J^{\rm E}=\sum_k j_k^{\rm E}$ however, is not). Because $J^{\rm E}_0$ has in addition a nonzero overlap with the magnetization current it can be used to bound the spin Drude weight away from zero. Let us denote by $J^{\rm S}=\sum_k j_k^{\rm tot}$ the total magnetization current. The thermodynamic overlaps needed, e.g., $\langle J^{\rm E}_0 J^{\rm S}\rangle$, are relatively simple to evaluate at infinite temperature but finite chemical potential, where the grand-canonical state is $\rho = \prod_k \rho_k$, with $\rho_k \sim \exp{( -\phi \sigma_k^{\rm z})}$ being the equilibrium state of one spin. Identifying $z={\rm tr}\,(\rho_k \sigma_k^{\rm z})$ as the average equilibrium magnetization, or, equivalently, as the filling fraction $f=(z+1)/2$, all averages are polynomial functions of $z$. Denoting by $D_{\rm S}$ the magnetization Drude weight, Mazur's inequality~\cite{Zotos:97} can be used to obtain
\begin{eqnarray}
D_{\rm S} &\ge& \frac{\beta}{2} K,\qquad K=\frac{1}{L}\frac{\langle J^{\rm E}_0 J^{\rm S}\rangle^2}{\langle J^{\rm E}_0J^{\rm E}_0\rangle},\\
K&=& 2^9\frac{z^2(1-z^2)(1+z^2)^2}{15+13z^2+10z^4+2z^6},\nonumber
\end{eqnarray}
holding at close-to infinite temperature. Observe that the bound is independent of $U$, simply because $J_0^{\rm E}$ and $J^{\rm S}$ are, even though the Hamiltonian in Eq.~(\ref{eq:Betheladder}) does depend on $U$. Away from maximal polarization, $z \neq \pm 1$, and nonzero magnetization, $z \neq 0$, the value of $K$ is nonzero, proving nonzero spin Drude weight in the integrable spin ladder (\ref{eq:Betheladder}) at an infinite temperature, and, as a consequence, ballistic magnetization transport. In the present work we shall numerically study the case $z=0$ ($f=1/2$) where there remains the possibility to have a non-ballistic transport. 

\subsubsection{Numerical results for zero-magnetization sector}

\begin{figure}[ht!]
\centerline{\includegraphics[angle=-90,width=0.45\textwidth]{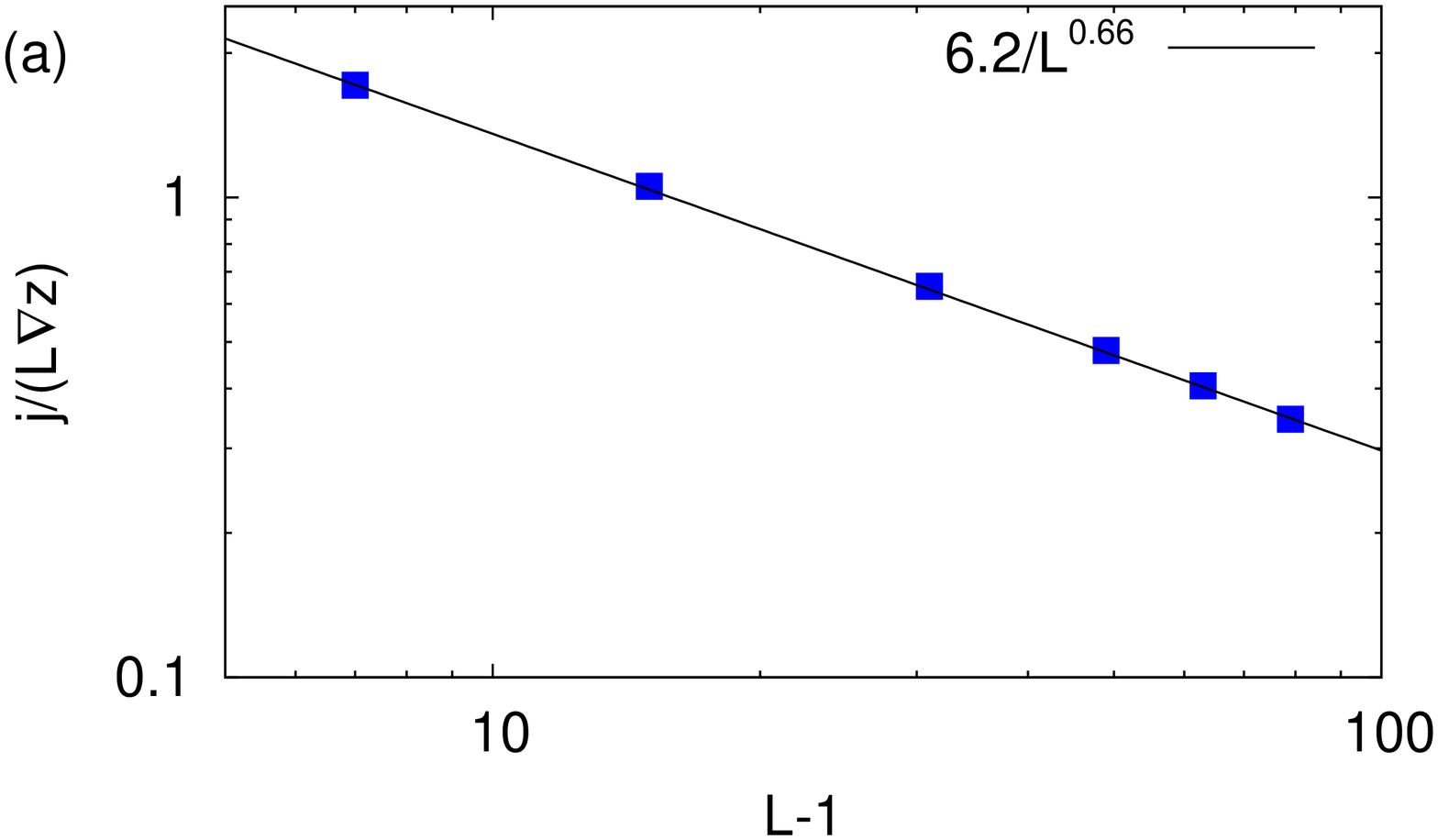}}
\centerline{\includegraphics[angle=-90,width=0.45\textwidth]{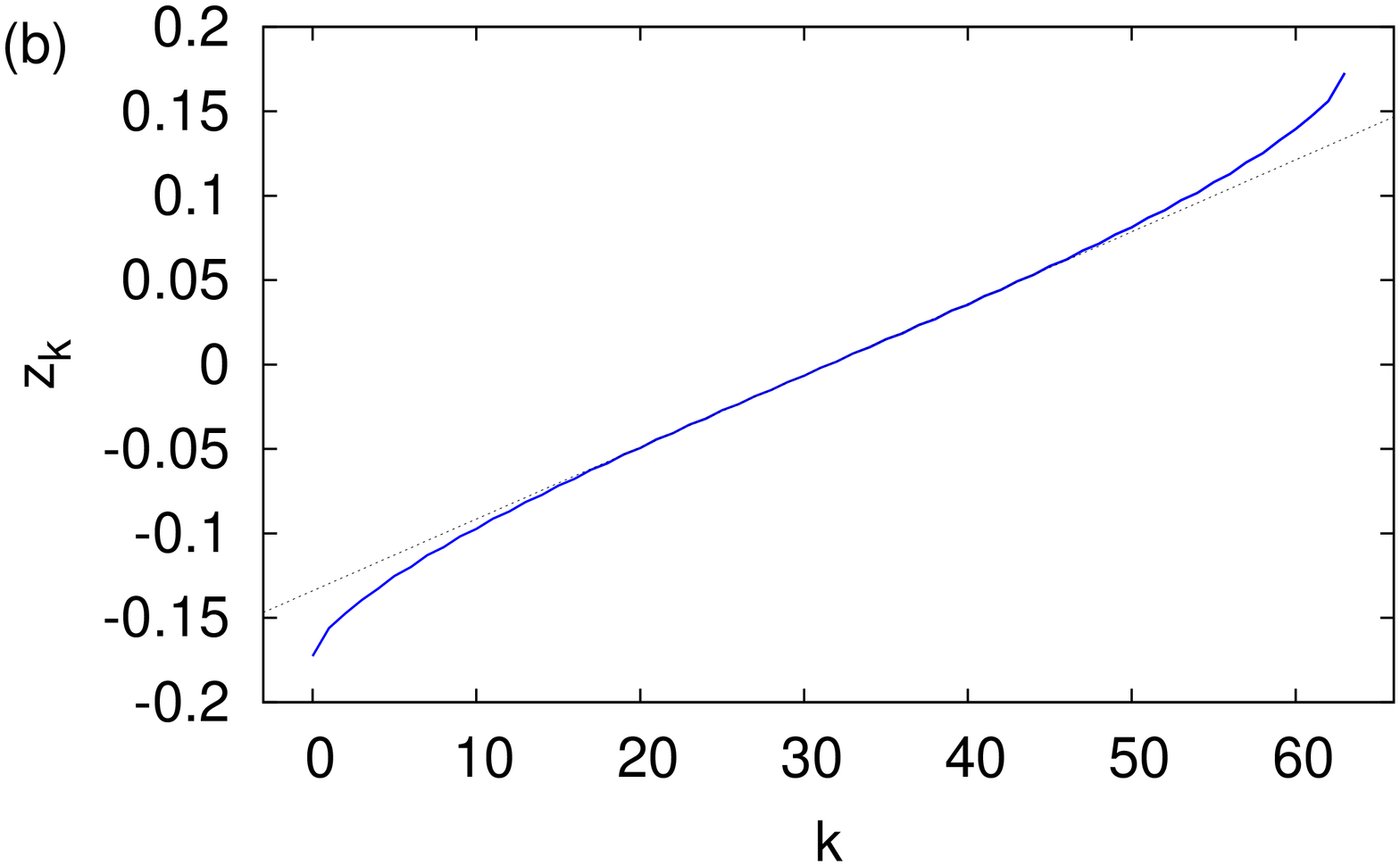}}
\caption{a) Dependence of the scaled magnetization current and b) magnetization profile in the upper leg ($L=64$) for the integrable ladder system, Eq.~(\ref{eq:Betheladder}). In b) the dotted line is the best-fitting gradient used in the scaling of the current shown in frame a). Parameters are $U=1$, and driving $\mu=0.2$.}
\label{fig:jB}
\end{figure}

In numerical simulations we shall use the critical $U=1$. Note however, that, due to the fact that the interaction term proportional to $U$ is equal to the constant of motion $C_3$ of the closed system, the magnetization transport for our symmetric driving is almost independent of the value of $U$. If the symmetry $V=\exp{(-\ii \alpha \sum_k \bm{\sigma}_k\cdot\bm{\tau}_k )}$ corresponding to the conserved quantity $C_3=\sum_k \bm{\sigma}_k\cdot\bm{\tau}_k$, would be an exact symmetry of the Liouvillian (\ref{eq:Lin}), $V {\cal L}(\rho)V^\dagger={\cal L}(V\rho V^\dagger)$, then the NESS state $\rho_\infty$ would be exactly independent of $U$. In our case the symmetry $V$ preserves the unitary part, $VHV^\dagger=H$, but is not an exact symmetry of the dissipative part (\ref{eq:ladderL}). Therefore, in an open system $V$ is only an approximate symmetry; it is violated at boundaries. Still, we find~\cite{foot2} that the magnetization transport is almost independent of $U$. This also shows that the size of the ground state gap by itself does not play any role on the transport at an infinite temperature.

In Fig.~\ref{fig:jB} we show the scaling of the magnetization current~\cite{foot1} of one leg species $j\equiv \langle j_k^\sigma (1+\bm{\tau}_k\cdot \bm{\tau}_{k+1}) \rangle$ with $L$ and one instance of the magnetization profile. The current scales as $j \sim 1/L^{0.66}$, indicating anomalous transport. Correspondingly, the magnetization profile along the ladder is not linear but rather displays larger gradients towards the ends. Similar profiles have been observed~\cite{Znidaric:11,Znidaric:11b} in the isotropic Heisenberg model, also showing anomalous transport $j \sim 1/L^\alpha$ with $\alpha=1/2$. Note that both, the isotropic Heisenberg model and the integrable ladder (\ref{eq:Betheladder}), are special due to their SU(2) and SU(4) symmetry, respectively. On a speculative note, considering that $\alpha=1/2$ for the SU(2) model, and $\alpha=2/3$ for the SU(4) one, the general rule would be that the exponent of anomalous transport is $\alpha=N/(1+N)$ for a permutation model $H\sim\sum_k P_{k,k+1}$ with an SU(2N) invariance. Because an SU(2N) model has $2N$ local levels, it could be written as a spin-$(N-\frac{1}{2})$ chain. Therefore, as $N \to \infty$ one goes essentially to the classical limit for which $\alpha \to 1$, i.e., one would get a diffusive transport. For a recent study of transport in the classical Heisenberg model see Ref.~\cite{Steinigeweg:12a}.

\section{Next-nearest-neighbor chains}

\begin{figure}[ht!]
\centerline{\includegraphics[width=0.25\textwidth]{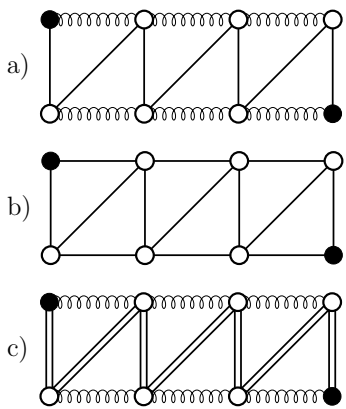}}
\caption{Different spin chains with a next-nearest-neighbor coupling: a) the XX chain with a ZZ n.n.n. coupling, b) the XX chain with an XX n.n.n. coupling, c) the isotropic Heisenberg chain with a ZZ n.n.n. coupling (a straight line is an XX-type coupling, a spring a ZZ-type coupling while a double line is an isotropic Heisenberg coupling). Full points mark the sites that are coupled to a reservoir described by Eq.~\ref{eq:chainL}.} 
\label{fig:verige}
\end{figure}

It is believed that integrability-breaking perturbations in 1d spin chains, provided they are large enough, will in general induce diffusive transport. This is expected on general grounds, because a sufficient perturbation will results in a chaotic system, and is also supported by numerical observations~\cite{Zotos:96,Rosch:00,Alvarez:02,Meisner:03,Mukerjee:08,Prosen:09,Steinigeweg:11,Steinigeweg:12,Huang:12}. At sufficiently low temperatures though, some studies~\cite{Fujimoto:03,Heidarian:07} observed indications of ballistic transport. In the present work we reconsider the question of magnetization transport in spin chains with integrability-breaking next-nearest-neighbor coupling in the regime of strong integrability-breaking (quantum chaos). The different spin chains studied are shown in Fig.~\ref{fig:verige}. Note that by numbering ladder sites in a zig-zag manner, the next-nearest-neighbor coupling of a chain is in a ladder formulation given by the coupling terms in both legs, while the nearest-neighbor coupling of a chain is a ladder coupling in rungs and the diagonal inter-rung coupling. 

\begin{figure}[ht!]
\centerline{\includegraphics[angle=-90,width=0.47\textwidth]{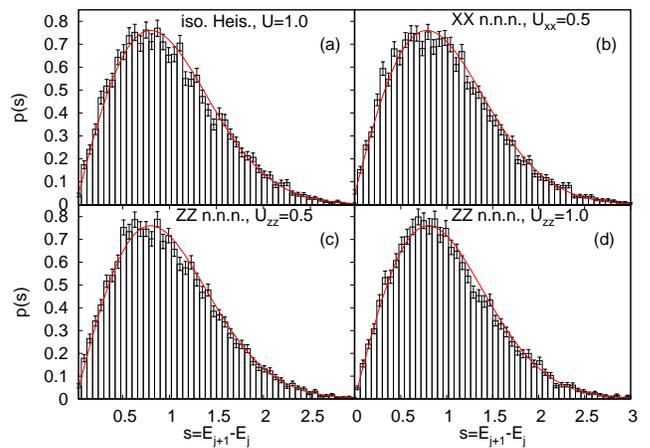}}
\caption{Level spacing distribution for spin chains. a) The isotropic Heisenberg with a ZZ n.n.n. coupling, b) the XX chain with a XX n.n.n. coupling, c) and d) the XX chain with a ZZ n.n.n. coupling. All data is for $L=8$ and a sector with $Z=0$ (averaging over 4 subsectors; in total around 11000 levels for each system), see the Appendix~\ref{app:symm}. The full curve is the Wigner's surmise (\ref{eq:Wigner}).}
\label{fig:lsdchain}
\end{figure}

\subsection{XX chain}
First, we shall study the XX chain with a ZZ next-nearest-neighbor coupling,
\begin{equation}
H=\sum_{i=1}^{L-1} (\sigma_i^{\rm x} \sigma_{i+1}^{\rm x}+\sigma_i^{\rm y} \sigma_{i+1}^{\rm y})+U_{\rm zz}\sum_{i=1}^{L-2} \sigma_i^{\rm z} \sigma_{i+2}^{\rm z}.
\label{eq:nnnchain}
\end{equation} 
The magnetization current in the NESS is a standard $j=\langle 2(\sigma_k^{\rm x}\sigma_{k+1}^{\rm y}-\sigma_k^{\rm y}\sigma_{k+1}^{\rm x})\rangle$. Integrability breaking perturbation of strength $U_{\rm zz}=0.5$ and $U_{\rm zz}=1.0$ shall be used, for which the system is quantum chaotic. In Fig.~\ref{fig:lsdchain} we can see a nice agreement of the level spacing distribution with Wigner's surmise.

\begin{figure}[ht!]
\centerline{\includegraphics[angle=-90,width=0.45\textwidth]{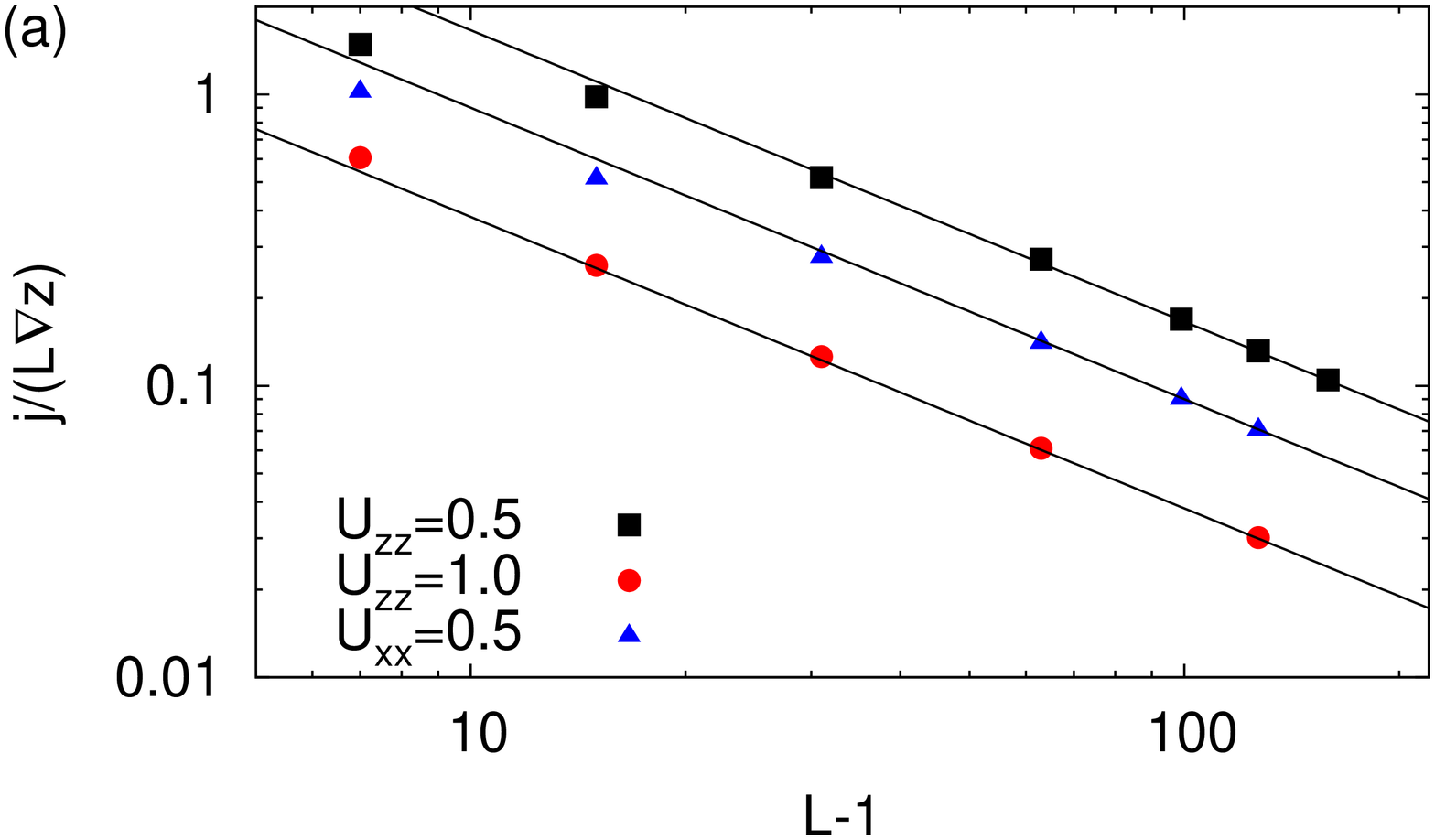}}
\centerline{\includegraphics[angle=-90,width=0.45\textwidth]{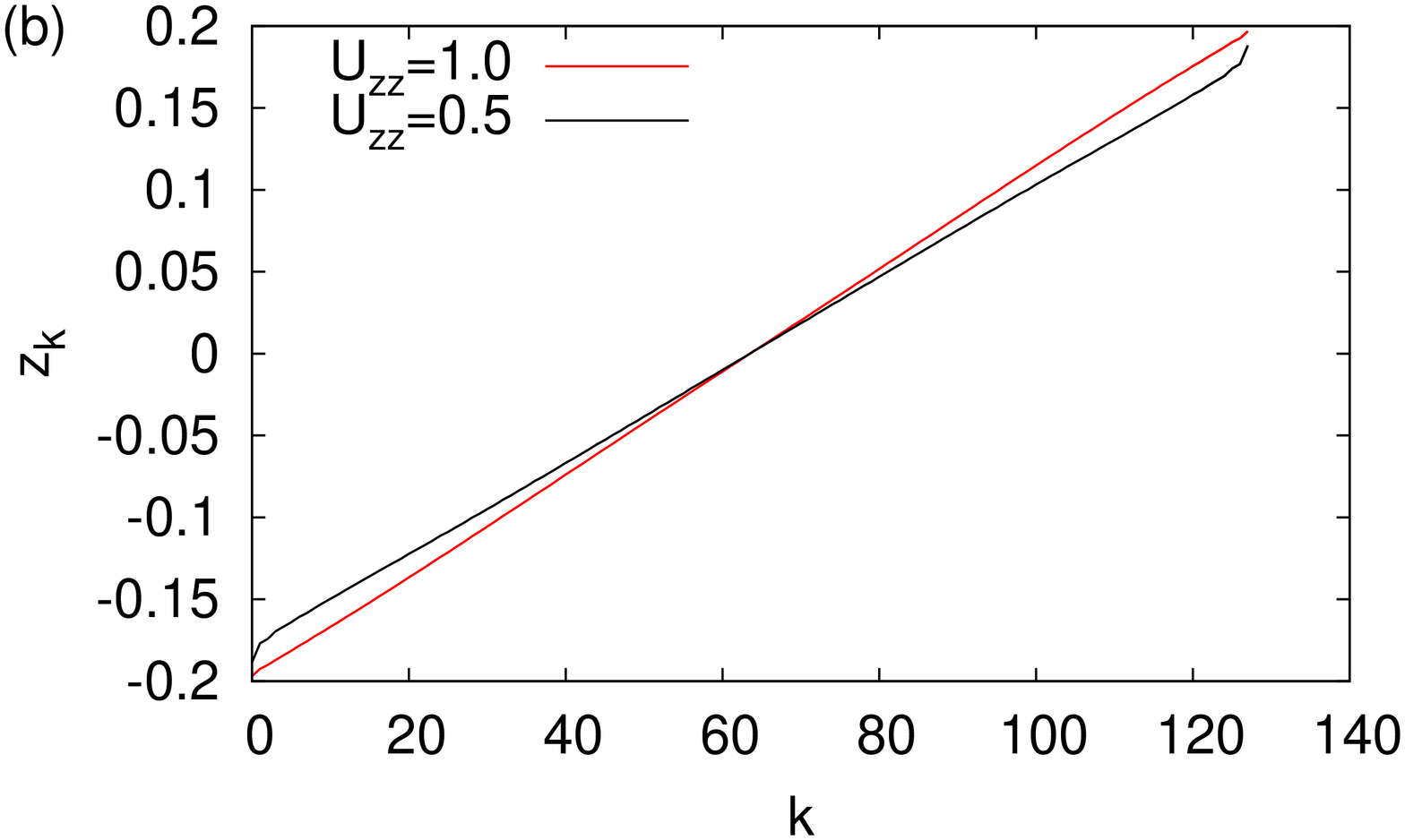}}
\caption{a) Scaling of the current for the XX chain with a ZZ (squares and circles) or an XX type n.n.n. coupling (triangles). The scaling is in all cases diffusive. Straight lines are $16.6/L$, $9.0/L$ and $3.8/L$ (top to bottom). b) The magnetization profile is linear ($L=128$). Driving is in all cases $\mu=0.2$.}
\label{fig:XXchains}
\end{figure}

We shall also study the XX chain with a XX type n.n.n. coupling,
\begin{equation}
H=\sum_{i=1}^{L-1} (\sigma_i^{\rm x} \sigma_{i+1}^{\rm x}+\sigma_i^{\rm y} \sigma_{i+1}^{\rm y})+U_{\rm xx}\sum_{i=1}^{L-2} \sigma_i^{\rm x} \sigma_{i+2}^{\rm x}+\sigma_i^{\rm y} \sigma_{i+2}^{\rm y}.
\label{eq:xxnnnchain}
\end{equation}
At the $U_{\rm xx}=0.5$ studied the model is again quantum chaotic, see Fig.~\ref{fig:lsdchain}. The magnetization current operator~\cite{footlast} gets in this case an additional next-nearest-neighbor term, and is 
\begin{equation}
j_k=2(\sigma_k^{\rm x} \sigma_{k+1}^{\rm y}-\sigma_k^{\rm y} \sigma_{k+1}^{\rm x})+2U_{\rm xx}(\sigma_k^{\rm x} \sigma_{k+2}^{\rm y}-\sigma_k^{\rm y} \sigma_{k+2}^{\rm x}).
\end{equation}
As one can see in Fig.~\ref{fig:XXchains} the magnetization transport is in all cases diffusive, indicated by the scaling $j \sim 1/L$, as well as by the linear magnetization profiles.

\subsection{Isotropic Heisenberg chain}
\begin{figure}[ht!]
\centerline{\includegraphics[angle=-90,width=0.45\textwidth]{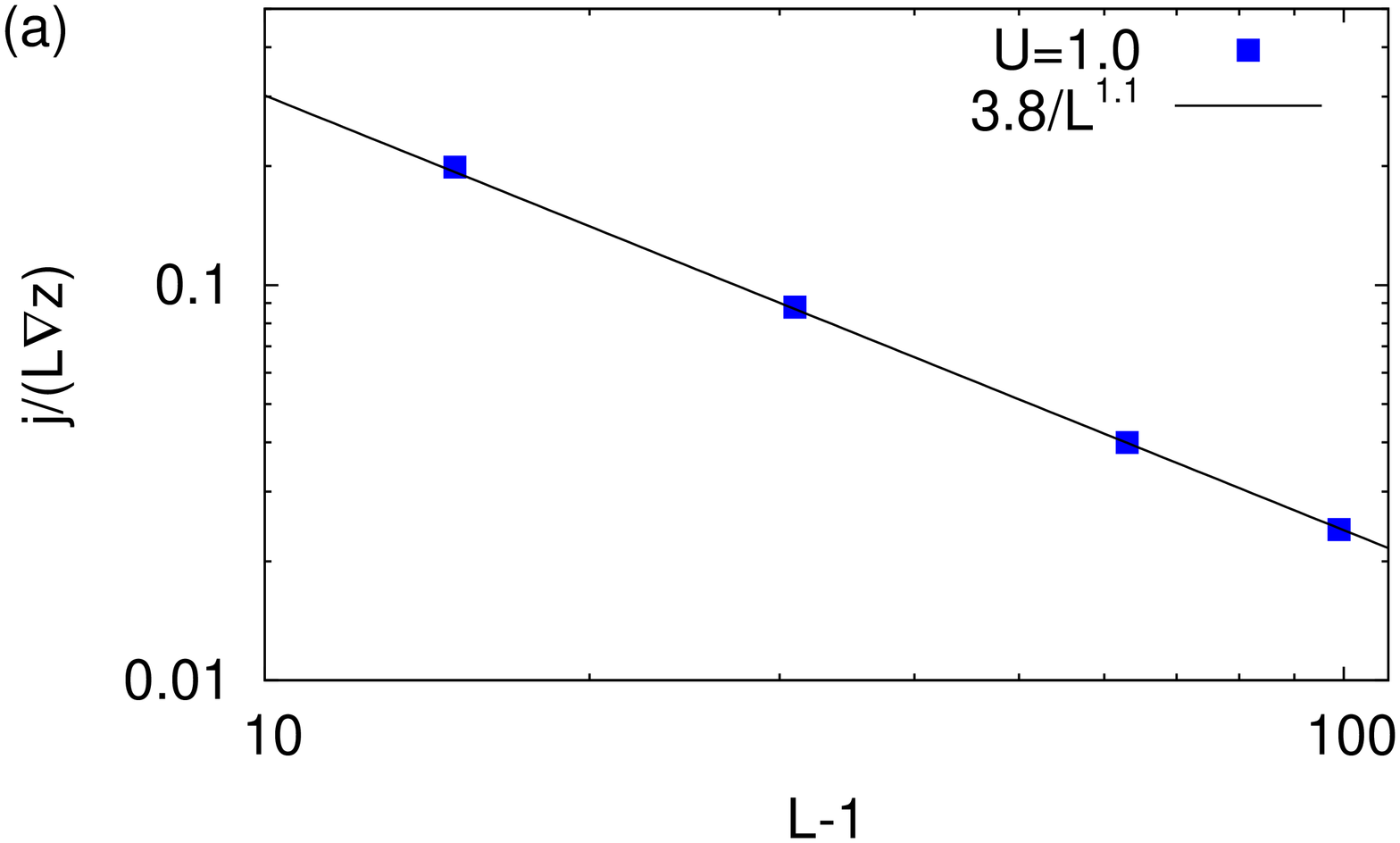}}
\centerline{\includegraphics[angle=-90,width=0.45\textwidth]{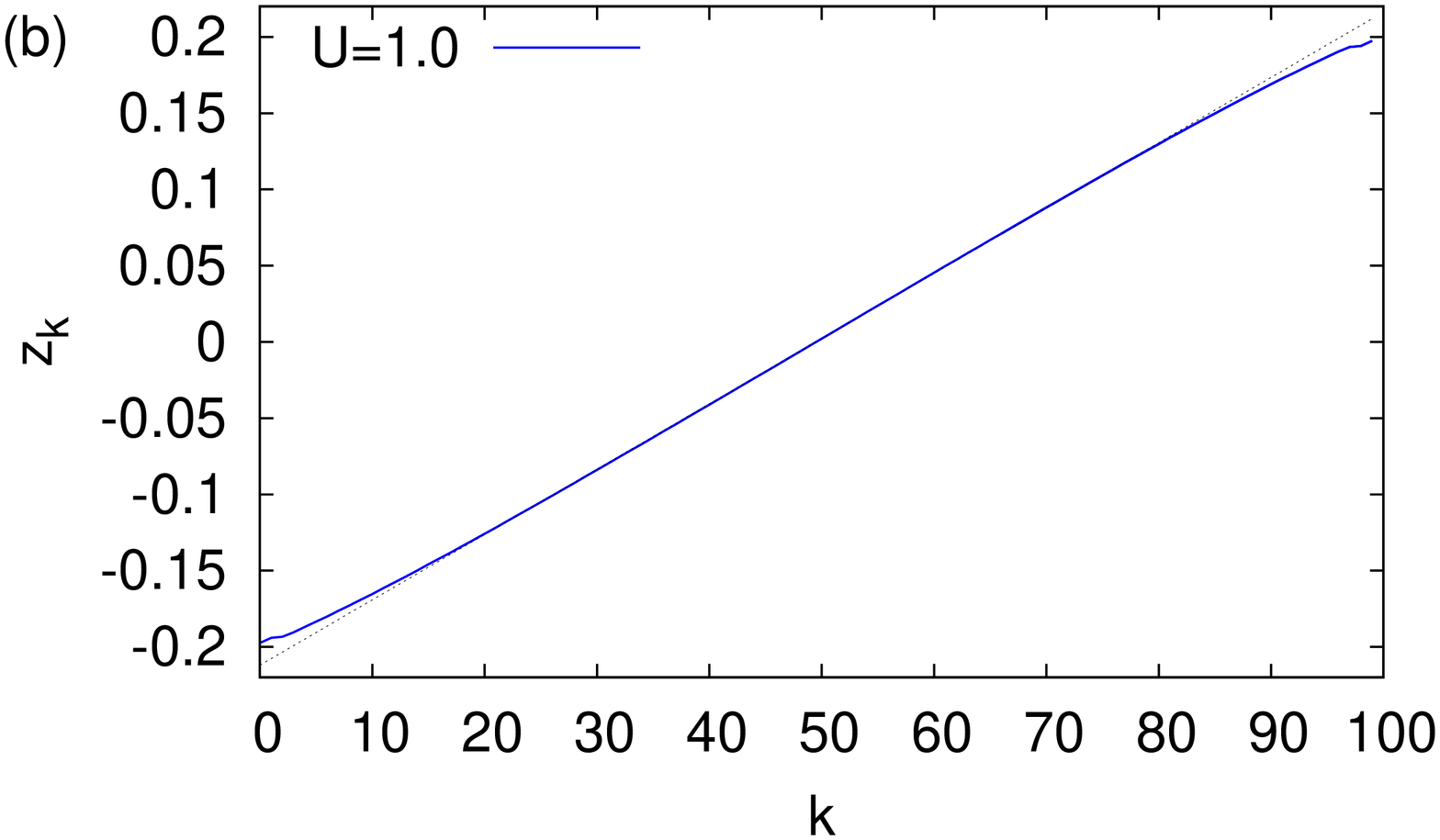}}
\caption{Isotropic Heisenberg chain with a ZZ n.n.n. coupling of strength $U=1.0$, $\mu=0.2$. a) scaling of the current with the system size, b) the magnetization profile for $L=100$.}
\label{fig:Isochains}
\end{figure}

As the last model we shall study the Isotropic Heisenberg chain with a ZZ n.n.n. coupling,
\begin{equation}
H=\sum_{i=1}^{L-1} \bm{\sigma}_i\cdot \bm{\sigma}_{i+1}+U\sum_{i=1}^{L-2} \sigma_i^{\rm z} \sigma_{i+2}^{\rm z},
\label{eq:IsoZnnnchain}
\end{equation} 
with $U=1.0$, for which the model is quantum chaotic, Fig.~\ref{fig:lsdchain}. The magnetization current is $j=\langle 2(\sigma_k^{\rm x}\sigma_{k+1}^{\rm y}-\sigma_k^{\rm y}\sigma_{k+1}^{\rm x})\rangle$. As shown in Fig.~\ref{fig:Isochains} the current scales as $j \sim 1/L^{1.1}$, while the profiles show slight deviations from a linear function close to the chain ends. Note that, as is most often the case, if the current scales faster than $\sim 1/L$, i.e., if the system goes towards an insulating regime, a local gradient in profiles is smaller close to the system edge (Fig.~\ref{fig:Isochains}), on the other hand, if the scaling is slower than $\sim 1/L$, i.e., if the system goes towards a ballistic regime, the gradient is larger (e.g., Fig.~\ref{fig:jB}). In the present case deviations are small and it is difficult to asses if it is just a finite size effect and the system becomes diffusive in the thermodynamic limit. 
Provided the scaling is asymptotically $\sim 1/L$, the found prefactor $3.8$ would be equal to the diffusion constant, $D \approx 3.8$. The same value~\cite{foot3} of the diffusion constant was found in Ref.~\onlinecite{Steinigeweg:11} using a current autocorrelation function obtained by an exact diagonalization as well as perturbatively for large $U$ via a time-convolutionless projection operator approach. 

\section{Conclusion}
We have studied magnetization transport in a linear response regime at an infinite temperature by numerically calculating nonequilibrium stationary states of the Lindblad master equation. For the isotropic Heisenberg ladder, being quantum chaotic, we find close-to diffusive behavior, with the differences being possibly due to finite-size effects. In the XX spin chain with strong next-nearest-neighbor interaction transport is always found to be diffusive. The isotropic Heisenberg chain with an integrability-breaking next-nearest-neighbor interaction is also very close to diffusive. We also found that the integrable ladder, which at $U=0$ has an SU(4) symmetry, shows anomalous magnetization transport in the zero-magnetization sector, while away from the zero-magnetization sector, using Mazur's inequality, we prove that the transport is ballistic.

\section*{Acknowledgments}

I acknowledge support by program P1-0044 of the Slovenian Research Agency.

\appendix

\section{Symmetries}
\label{app:symm}
We study magnetization transport, i.e., transport of the $z$-component of spin. All systems considered (their hamiltonian part) therefore conserve the total magnetization in the $z$-direction. For ladders this is the operator $Z=\sum_{j=1}^L \sigma_k^{\rm z}+\tau_k^{\rm z}$. The corresponding symmetry transformation is a rotation $U_{\rm z}=\prod_j \exp{(-\ii \alpha \sigma_j^{\rm z})}\exp{(-\ii \alpha \tau_j^{\rm z})}$. Under $U$ Pauli matrices transform as $U\sigma^{\rm z}_k U^\dagger = \sigma_k^{\rm z}$, $U\sigma_k^{\rm x} U^\dagger = \cos{(2\alpha)}\sigma_k^{\rm x}+\sin{(2\alpha)}\sigma_k^{\rm y}$, and $U \sigma_k^{\rm y} U^\dagger = -\sin{(2\alpha)}\sigma_k^{\rm x}+\cos{(2\alpha)}\sigma_k^{\rm y}$, and similarly for $\tau_k^\alpha$.

There are two geometrical symmetries of the underlying ladder lattice. One is a parity $P_{\rm x}$ in the $x$-direction, obtained by mapping of sites $k \to L+1-k$, while the other is a parity $P_{\rm y}$ in the $y$-direction, obtained by the swapping of the two legs, $\sigma_k^\alpha \leftrightarrow \tau_k^\alpha$.

In addition, there is a spin-flip symmetry given by the transformation $U=\prod_j \sigma_j^{\rm x} \tau_j^{\rm x}$, i.e., a rotation $\exp{(-\ii \pi \sigma^{\rm x}/2)}$ around the $x$-axis. It changes the sign of $\sigma_k^{\rm y,z}$ while it preserves $\sigma_k^{\rm x}$. It commutes with the rotation $U_{\rm z}$ around $z$ only in the sector with zero total magnetization $Z=0$.

Symmetries of the isotropic Heisenberg ladder described by $H$ in Eq.~(\ref{eq:Heisladder}) are both parities $P_{\rm x}$ and $P_{\rm y}$, spin-flip and total magnetization $Z$. In addition, the square of the total spin $(\sum_j \bm{\sigma}_j+\bm{\tau}_j)^2$ is also a constant of motion.

For chains with a n.n.n. coupling, Eqs.~(\ref{eq:nnnchain},\ref{eq:xxnnnchain},\ref{eq:IsoZnnnchain}), the symmetries are spin-flip, total magnetization $Z$ and the product of parities $P_{\rm x}P_{\rm y}$.


\begin{thebibliography}{99}
\bibitem{footE} We have also numerically checked that at $\mu=0.2$ the expectation values of the energy density in the NESS are indeed small ($\propto \mu^2$) and the states are therefore at inverse temperatures close to zero.
\bibitem{foot0} $\mu=0.2$ seems to be a good choice regarding numerical efficiency, while still being at the upper end of the linear response regime.
\bibitem{foot2} We have checked numerical simulations at $U=0.3$ and $U=0.0$ and they resulted in virtually the same $j(L)$ as for $U=1.0$.
\bibitem{foot1} Numerical simulations for the integrable ladder are somewhat more demanding than for other systems studied because the required matrix product operator dimension $M$ is larger. Because of slower convergence with $M$ we have used extrapolation in order to obtain the current $j$ in the limit $M \to \infty$. That is, at each $L$ we obtain current $j(M)$ from simulations with different $M$, up-to the largest one $M_{\rm max}$. We then try to use polynomial extrapolation, fitting to $j(M)$, and trying to extrapolate the current at $M \to \infty$. The error is then estimated as the difference between the extrapolated value and $j(M_{\rm max})$. For instance, in the worst case of $L=80$, for which $M_{\rm max}=150$, the difference was about $5\,\%$.
\bibitem{footlast} Using the symmetric current definition $j_k+2U_{\rm xx}(\sigma_{k-1}^{\rm x} \sigma_{k+1}^{\rm y}-\sigma_{k-1}^{\rm y} \sigma_{k+1}^{\rm x})$ has very little influence on the results presented. The scaling is in such case $j/(L\nabla z)\approx 10/L$. I would like to thank M.~Medenjak for pointing this out.
\bibitem{foot3} Comparing the diffusion constant $D$ in Pauli variables (Eq.~\ref{eq:Four}) with the diffusion constant $D_{\rm spin}$ in spin variables $s^\alpha=\frac{1}{2}\sigma^\alpha$, there is a factor of $4$ between the two, $D=4D_{\rm spin}$.
\end{thebibliography}
\end{document}